\documentclass[aps,prl,twocolumn,superscriptaddress]{revtex4}
\usepackage{amsmath}
\usepackage{graphicx}
\begin{document}
\title{Quantum Monte Carlo study of the three-dimensional attractive Hubbard
model}
\author{Alain Sewer}
\affiliation{IRRMA, EPFL, 1015 Lausanne, Switzerland}
\affiliation{Institut de Physique, Universit\'e de Neuch\^atel, 
2000 Neuch\^atel, Switzerland}
\author{Xenophon Zotos}
\affiliation{IRRMA, EPFL, 1015 Lausanne, Switzerland}
\author{Hans Beck}
\affiliation{Institut de Physique, Universit\'e de Neuch\^atel, 
2000 Neuch\^atel, Switzerland}
\date{\today}

\begin{abstract}
We study the three-dimensional (3D) attractive Hubbard model by means of the
Determinant Quantum Monte Carlo method. This model is a prototype for  the
description of the smooth crossover between BCS superconductivity and 
Bose-Einstein condensation. By detailed finite-size scaling  we extract the
finite-temperature phase diagram of the model. In particular, we interpret the
observed behavior according to a scenario of two fundamental temperature 
scales; $T^*$ associated with Cooper pair formation and $T_c$ with 
condensation (giving rise to long-range superconducting order). Our results
also indicate the presence of a recently conjectured phase  transition hidden
by the superconducting state. A comparison with the 2D case is briefly
discussed, given its relevance for the physics of high-$T_c$ cuprate
superconductors. 
\end{abstract}


\maketitle

\begin{figure}
\includegraphics[height=5.0 cm]{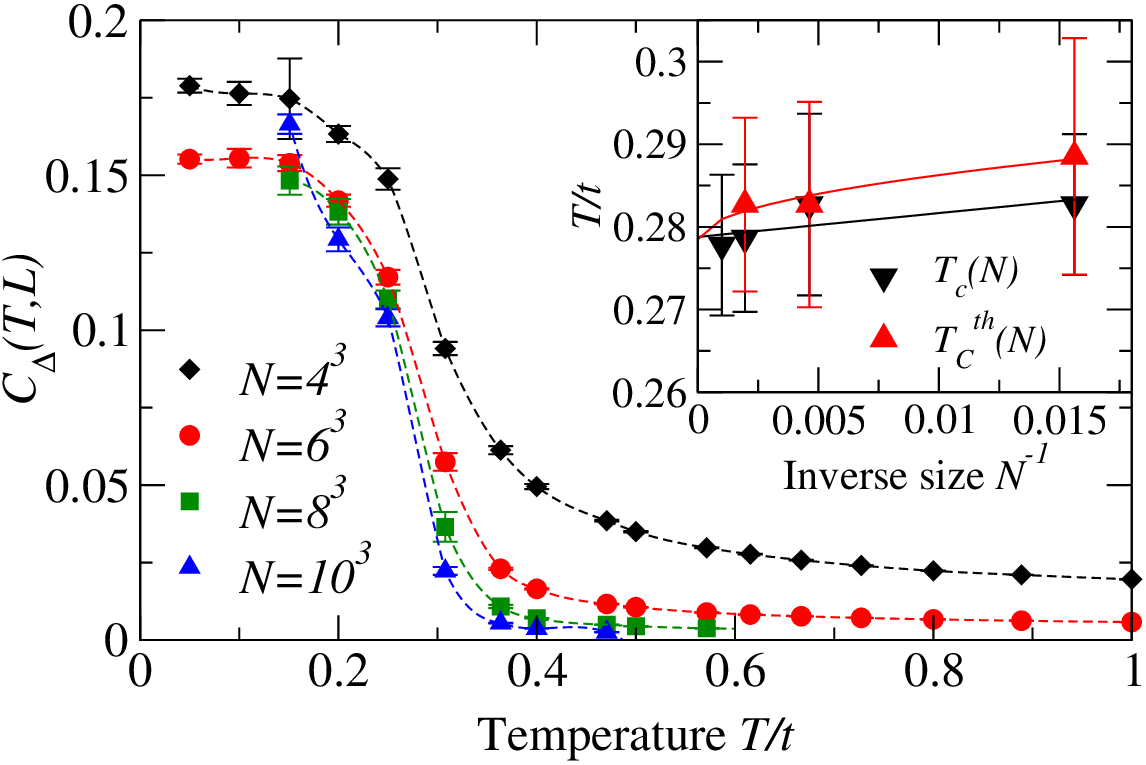}
\caption{{\it Main}: temperature and size dependences of the pair-pair
correlation function (\ref{e2}) for the case $U=6t$ and $n=0.5$. {\it Inset}:
extrapolation to the thermodynamic limit of the size-dependent critical
temperatures $T_c(N)$ and $T_c^{th}(N)$, same $U$ and $n$.   
\label{FIG1}}
\end{figure}

\begin{figure}
\includegraphics[height=5.0 cm]{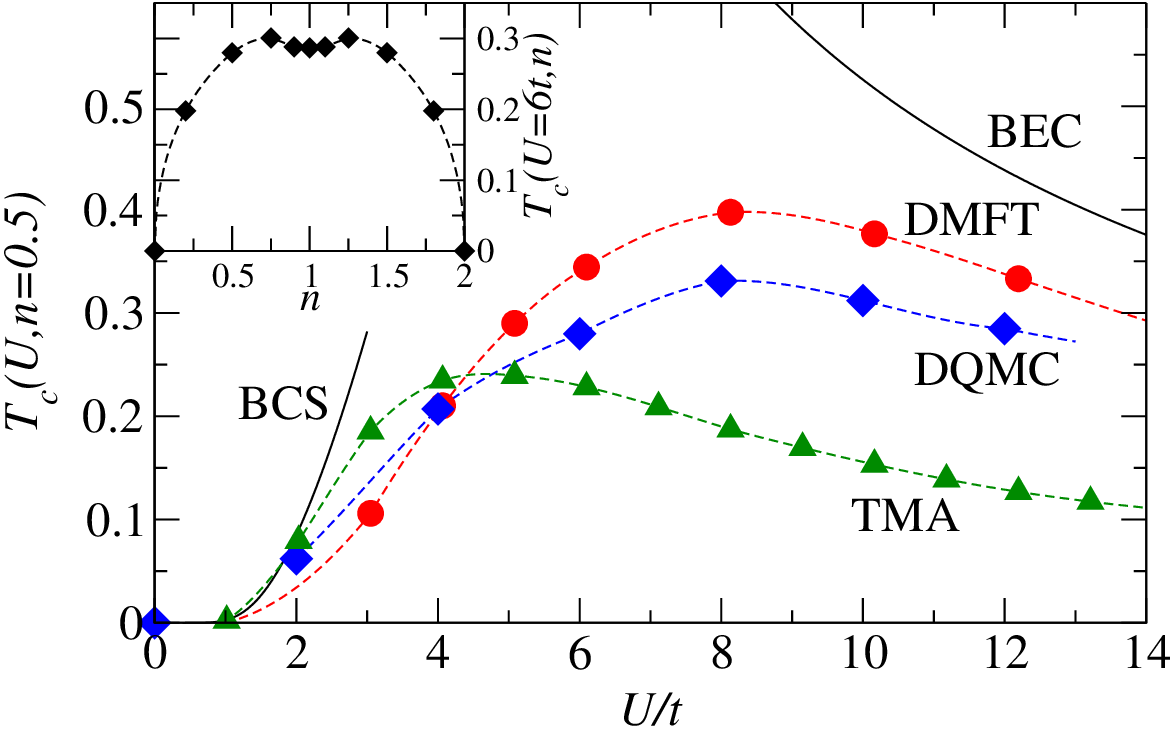}
\caption{DQMC results for the critical temperature $T_c(U,n)$ and comparison
with other methods \cite{keller}. {\it Main}: dependence on the coupling $U$.
{\it Inset}: dependence on the density $n$.   
\label{FIG2}}
\end{figure}

\begin{figure}
\includegraphics[height=7.0 cm]{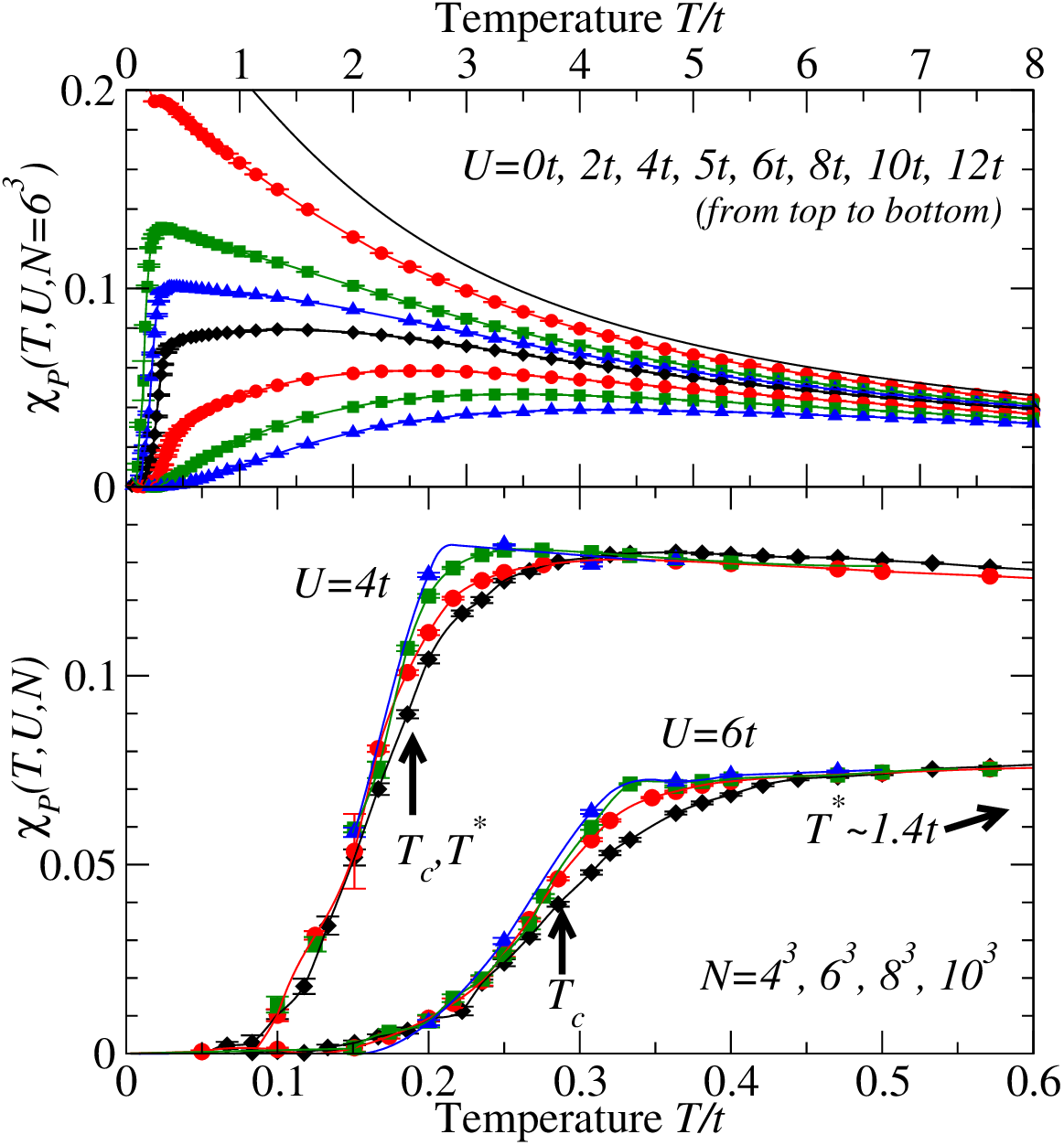}
\caption{Pauli susceptibility $\chi_P$. {\it Top}: $T$ dependence for various
values of $U$ (size $N=6^3$). {\it Bottom}: $T$ and $N$ dependence close to the
transition temperature and separation of $T_c$ and $T^*$, same symbols as in
Fig. \ref{FIG1} ($n=0.5$ for both cases).
\label{FIG3}}
\end{figure}

\begin{figure}
\includegraphics[height=4.0 cm]{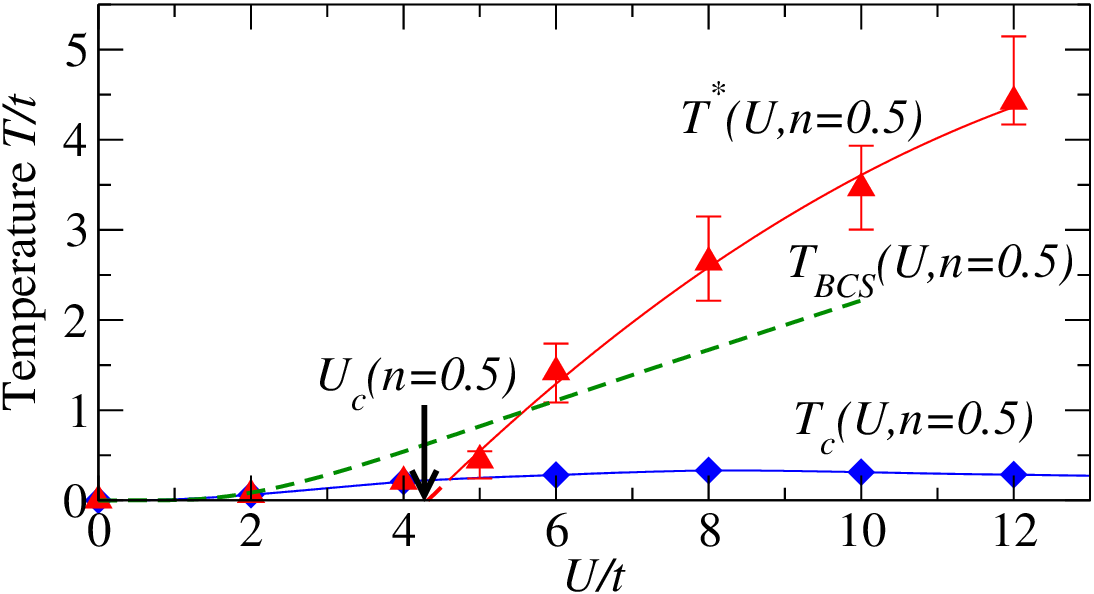}
\caption{$U$-$T$ phase diagram of the 3D attractive Hubbard model at
quarter-filling. The errorbars correspond to the temperature interval around
$T^*$ where $\chi_P(T)$ is less than one percent smaller than its maximal
value, giving thereby an idea of the temperature range associated with
$T^*$.    
\label{FIG4}}
\end{figure}

The existence of a smooth crossover between the two paradigms of quantum
superfluidity, the Bardeen-Cooper-Schrieffer (BCS) superconductivity and the
Bose-Einstein condensation (BEC) is firmly established  \cite{legett,nsr}. In
this context, the attractive Hubbard model (AHM) has appeared as an ideal
presentation of the  whole evolution between the BCS and BEC physics
\cite{randeria}. A concrete property of this Hamiltonian is the existence of
two (not always) distinct energy scales: one associated with the formation of
Cooper pairs ($T^*$) and another with the onset of long-range order in the
system ($T_c$) \cite{ranninger}. Although their qualitative behavior is
well-known, a quantitative determination is still missing, due to the fact that
it is hard  to access the intermediate regime by a controlled approximation
scheme. In this respect the Determinant Quantum Monte Carlo (DQMC) method
\cite{hirsch,lohgub} is a powerful tool as it provides  results free of
systematic errors. A detailed finite-size analysis is however necessary in
order to extract the thermodynamic limit  properties, which can then be
compared with the outputs of other methods recently applied to the same problem
\cite{keller,capone}. At this point we should stress the role of dimensionality
that determines the nature of the superconducting phase transition at $T_c$;
the strictly 2D realization of the model is characterized by a
Berezinskii-Kosterlitz-Thouless type phase  transition, whereas the 3D case
displays a ``normal'' second-order one, which is more easily accessible by
DQMC.  Since the intermediate regime of the AHM constitutes the simplest model
for  a short-coherence-length superconductor, the considerations presented 
hereafter may as well help to clarify the influence of the dimensionality on 
some properties exhibited by the 3D strongly anisotropic high-$T_c$
superconductors.
\\
In this Letter, we present the results of extensive DQMC simulations for the
finite-temperature properties of the AHM in three dimensions. In spite of
finite-size effects, we show that it is possible by a scaling analysis to
quantitatively establish the  phase diagram of $T_c(U,n)$ as a function of the
interaction strength and density of a model that exhibits a genuine
second-order phase transition (unlike its 2D version). Furthermore,  the pair
formation temperature $T^*$ is studied in detail, revealing the existence of a
transition in the non-superconducting state taking place at a critical coupling
strength. These results complete recent calculations which have postulated the
existence of such a transition in the infinite-dimension version of the model
\cite{keller,capone}. 
\\
{\it Model and method.} ---
The attractive Hubbard model is defined by the following Hamiltonian,
\begin{equation}
\mathcal{H}=
-t\sum_{<i,j>,\sigma}(c_{i\sigma}^{\dagger}c_{j\sigma}+\mbox{H.c.})
-U\sum_i n_{i\uparrow}n_{i\downarrow}
-\mu \sum_{i,\sigma}n_{i,\sigma},
\label{H}
\end{equation}
where $<i,j>$ denotes a pair of nearest neighbors on a cubic lattice with 
$N=L^3$ sites, $c_{i\sigma}^{\dagger} (c_{i\sigma})$ is a fermion creation 
(annihilation) operator of spin $\sigma=\uparrow\downarrow$ and 
$n_{i,\sigma}=c_{i\sigma}^{\dagger} c_{i\sigma}$. We take   $t>0$, $U>0$ and
the chemical potential $\mu$ is tuned to yield a fixed  density $0<n<2$. 
Outside half-filling ($n\neq 1$) this model presents a finite-temperature 
transition into a phase characterized by long-range $s$-wave superconducting
order associated with the breaking of the U(1) gauge symmetry.   
\\
To study the finite temperature properties of this system we use the
conventional DQMC \cite{hirsch,lohgub} simulation method.  Since the attractive
interaction does not lead to a minus-sign problem, the whole $U$-$n$-$T$ phase
diagram can be reliably studied. Because of the grand-canonical nature of DQMC,
it is necessary to estimate the function $\mu=\mu(T,n,U,L)$ in order to work at
a fixed density $n$.  This presents a considerable load in this work compared
to similar DQMC simulations at half-filling \cite{staudt}.  Typically we take
$n=0.5$ (quarter-filling) for which results using other  methods have already
been presented \cite{keller,capone}. We also restrict ourselves to 
finite-temperature static correlation functions \cite{lohgub}, such as the
$s$-wave pair-pair correlation function $C_{\Delta}$ and the Pauli spin
susceptibility $\chi_P$, given by: 
\begin{eqnarray}
C_{\Delta}(T,N)&=&\frac{1}{N}\sum_{i,j}
\langle\Delta_i\Delta_j^{\dagger}+\Delta_j\Delta_i^{\dagger}\rangle
\label{e2}\\
\chi_P(T,N)&=&\frac{1}{T}\,\frac{1}{N}\sum_{i,j}
\langle\mathbf{S}_i\cdot\mathbf{S}_j\rangle.
\label{e3}
\end{eqnarray}
Here $\Delta_i=c_{i\uparrow}c_{i\downarrow}$ and 
$\mathbf{S}_i=\sum_{\mu,\nu=\uparrow,\downarrow}
c_{i\mu}^{\dagger}\boldsymbol{\sigma}_{\!\mu\nu}c_{i\nu}$, 
$\boldsymbol{\sigma}$ being the vector of Pauli matrices. $C_{\Delta}$ allows
to determine the superconductiong transition temperature $T_c$, since  it
signals the breaking of the U(1) gauge symmetry. We recall that this approach
is not applicable to the strictly 2D case where more sophisticated quantities
have to be calculated \cite{fakher}. On the other hand $\chi_P$ indicates the
presence of pairing in the system, related to the temperature  scale $T^*$ as
discussed below. 
\\
Regarding the DQMC simulations, the imaginary time discretization is
$\Delta\tau=0.125 t^{-1}$ and  lattices of size $N=4^3 - 10^3$  (with periodic
boundary conditions) are  considered in order to keep the CPU time into
reasonable limits.  Two types of finite-size effects are present:  first, the
discreteness of the spectrum introduces artificial features at low temperatures
$T\lesssim 0.1t$ and weak couplings $U\lesssim 2t$ (signaled by
$(\frac{\partial\mu}{\partial T})_n>0$); second, the superconducting phase
transition is rounded and corresponds to  the point where the correlation 
length $\xi(T)$ becomes larger than the linear system size $L$.  
\\
{\it Determination of $T_c$.} ---
Extracted by  finite-size analysis of very good quality data, the value of
$T_c$ is in principle  free of systematic errors, except a small uncertainty 
($\lesssim 5$ percents) due to the statistical error and to the finite
imaginary time discretization $\Delta\tau>0$. Given $U$ and $n$, the pair-pair
correlation function $C_{\Delta}$ (Eq.\ref{e2}) is  evaluated for various
temperatures $T$ and sizes $N$. This shows clearly that $C_{\Delta}$ is
characterized by a low- and a high-temperature regime, related by a transition
region that becomes sharper and sharper as $N$ increases. The latter
observation agrees well with the behavior expected in the thermodynamic limit,
where $C_{\Delta}$ displays a discontinuous  derivative at the phase transition
and becomes non-zero only below $T_c$.  This behavior, typical for all the
parameter values used in our calculations,  is shown in Fig. \ref{FIG1} for the
special case $U=6t$ and $n=0.5$. 
Although it does not correspond to a genuine phase transition, it allows to
define a size-dependent transition temperature $T_c(N)$ which we can use to
deduce the value of $T_c\equiv T_c(\infty)$. A convenient choice for $T_c(N)$
is given by the inflexion  point of the curve $C_{\Delta}(T,N)$ versus $T$
obtained by a (stable)  Lagrange polynomial interpolation of the DQMC data.
Plotting the obtained $T_c(N)$ versus $N^{-1}$, we extrapolate to $N\rightarrow
\infty$  using a linear fit of the data, as shown in the inset of  Fig. 
\ref{FIG1}. The validity of this procedure is confirmed by  the evaluation of
the specific heat $c_V(T)$ whose well-defined peak can be used to define
another size-dependent critical temperature $T_c^{th}(N)$. $c_V(T)$ is obtained
by the  numerical derivative of the expectation value of the energy
\cite{thesis}. A fit of the values for $T_c^{th}(N)$ with a functional form
$T_c(\infty)=T_c^{th}(N)+\mathcal{O}(\frac{1}{\sqrt{N}})$ (corresponding to a
superconducting phase transition in the universality  class of the 3D $XY$
model \cite{engelbrecht,schneider})  is shown on the inset of Fig. \ref{FIG1}. 
It reveals that the finite-size corrections to $T_c$ are very weak and in
particular not larger than the statistical errors resulting from the DQMC
method. Thus both approaches presented above are fully compatible and yield an
uncertainty on the extrapolated value of $T_c$ which is typically of the order
of 5 percent. 
\\
{\it The critical temperature} $T_c(U,n)$. ---
The above procedure, applied to a range of parameters $U$ and $n$,   determines
quantitatively the $U$-$n$-$T_c$ phase diagram of the AHM. First we consider
half-filling (i.e. $n=1$) which provides a useful check for our method. This
case is equivalent to the repulsive model that has been  recently studied by
Staudt {\it et al.} using the same method \cite{staudt}. The agreement on the
values of $T_c(U,n=1)$ is almost perfect \cite{thesis}; a small difference
($<3$ percents) appearing systematically is due to the extrapolation
$\Delta\tau\rightarrow 0$ performed by these authors and not done here due to 
calculation time restrictions. Turning now to quarter-filling, we obtain  the
results presented on Fig. \ref{FIG2}. Before discussing the intermediate  $U$
regime, we observe that, 
as long as the DMQC method works properly ($2t\le U\le 12t$), the extreme
values of $T_c(U)$ are joining progressively the corresponding asymptotic 
curves, given by the BCS gap equation for small $U$ and by the 3D BEC formula
for large $U$. Their respective dependences in $U$ follow essentially
$T_c\propto\exp(1/U)$ and $T_c\propto 1/U$, with the assumption that in the
latter case the bosons are noninteracting and have an effective hopping
amplitude $t_B=2t^2/U$. In the crossover region we observe, as expected, a
smooth interpolation between the BCS and BEC regimes, with a maximal value of
$T_c=0.35t$ situated at $U\simeq 8t$. It is now interesting to compare our
results with those proposed in recent works. In Fig. \ref{FIG2} the data
obtained using the Dynamical Mean-Field Theory (DMFT) and a 
$\mathbf{k}$-independent $T$-matrix approximation (TMA) \cite{keller} are also
plotted,  rescaled by a factor $\sim 2$ so that the dimensionless product $U$
times the density of states at the  Fermi level is the same as in our model
\cite{thesis,dos}.  Similarly to the half-filling case \cite{staudt}, we
observe  a good overall agreement with DMFT results, the discrepancy at strong
coupling ($U>6t$) being attributed to the mean-field character of DMFT; on the
other hand TMA clearly fails outside the BCS regime.  We also mention a recent
$\mathbf{k}$-dependent $T$-matrix calculation \cite{engelbrecht} for  $U=4t$
with a $T_c$ in quantitative agreement with our results.
\\
In addition to $U$, $T_c$ also depends upon the density $n$.  Our results show
that the function $T_c(U=\mbox{const.},n)$ is not monotonic \cite{dossantos} in
$0\le n \le 1$ ($1\le n \le 2$) unlike it was previously assumed
\cite{ranninger}. The maximal transition temperature for a given $U$ is
situated around $n=0.75$ ($1.25$).  This feature is illustrated on Fig.
\ref{FIG2} for the case $U=6t$ and is reminiscent of the two-dimensional case
where the higher symmetry of the Hamiltonian (\ref{H}) at half-filling (SO(3)
instead of U(1)) reduces $T_c$ to zero, as discussed recently \cite{tremblay}.
The appearance of an additional charge-density wave (CDW) ordering has been
studied by means of the  corresponding static correlation function
\cite{thesis}. 
\\
{\it The pairing temperature scale $T^*(U,n)$.}---
As mentioned previously, $T^*$ is besides $T_c$ another temperature scale that
characterizes the BCS-BEC crossover.  In the case of the AHM, $T^*$ can be
interpreted within a pairing scenario as signaling a  re-arrangement of
fermionic quasiparticles into $s$-wave singlet pairs. As a consequence, the
spectral weight of low-energy  spin excitations is reduced and the spin
response weakens. This process can be studied by considering the Pauli
susceptibility $\chi_P$  (Eq. \ref{e3}). Although $T^*$ may not always
correspond to a single point, but to an extended energy scale, it can
nevertheless be identified with the position of the maximum of $\chi_P(T)$
\cite{dossantos}.  This definition has the advantage of satisfying the expected
asymptotic behavior of $T^*$, i.e. $T^*=T_c$ in the BCS case and $T^*\propto
U/\ln(U/\epsilon_F)^{3/2}$ for the BEC limit  \cite{randeria}. The way the
Pauli spin susceptibility evolves between these  two regimes is shown on Fig.
\ref{FIG3}.
It is instructive to analyze these DQMC results by considering the sensitivity
of $T^*$ to finite-size effects. For the ``weak coupling'' case $U\le 4t$, one
observes that the shape of $\chi_P(T)$ in the region around its maximal value
depends strongly on the system size $N$, becoming sharper as $N$ increases. In
this case the extracted value of $T^*$ turns out to be nearly equal to $T_c$,
given the accuracy on the numerical results ($\lesssim 5$ percents).  On the
other hand, a ``strong coupling'' behavior appears for $U\ge 5t$, characterized
by a much smoother susceptibility around its maximum. In this region  
finite-size effects have disappeared, indicative of an effect characterized by
a short coherence length. Here, $T^*$ is definitely different from $T_c$.  In
the interval $[T_c,T^*]$ the interesting phenomenon of \emph{precursor pairing}
takes place, a point which will be further discussed below. We can thus present
the complete phase diagram on Fig.\ref{FIG4} by adding the function
$T^*(U,n=\text{const.})$ 
that clearly displays the two different regimes described above.  In the weak
coupling regime one observes that $T^*$ does not correspond to a BCS  critical
temperature extrapolated at $U\ge2t$. On the strong coupling side $T^*$ defines
an energy scale, which is  approximately quantified by the errobars on
Fig.\ref{FIG4}, and resembles to a straight line situated below the diagonal,
in qualitative agreement with the asymptotic expression given above. 
\\
{\it Discussion.}---
A first remark concerns the recent observation of a (first-order) phase
transition in the \emph{non-superconducting} solution of the AHM
\cite{keller,capone}. Since this state is metastable below $T_c$, it cannot be
accessed by DQMC (applying a magnetic field would cause a minus-sign problem).
However the manifestations of this transition may be present above $T_c$ as
well and the previous analysis of the Pauli spin susceptibility constitutes an
ideal illustration. Indeed, it turns out that the high-temperature behavior of
$\chi_P(T)$, observed for $U\le 4t$ and characterized by a \emph{monotonic}
decrease with $T$, may   correspond to a Fermi liquid normal state  where the
interaction amounts only to a renormalization of parameters. On the other hand,
the regime $U\ge 5t$, which displays the phenomenon of precursor pairing for
$T_c<T\lesssim T^*$, fits well to a phase containing ``incoherent pairs''
\cite{keller,capone}. Consequently a ``critical'' coupling strength $U_c$ may
be situated around $U=(4.3\pm0.1)t$, as it can be deduced by extrapolation at
$T=0$ on Fig. \ref{FIG4}. This value argrees very well with the (rescaled) DMFT
result $0.56\times W\times 2\approx 4.5t$, $W=4t$ being the bandwidth
\cite{keller,dos}. One also remarks that $U_c$ does not correspond to the point
where the chemical potential $\mu$ (including the Hartree shift $-U/2$) becomes
lower than the bottom of the non-interacting band (for $n=0.5$, we would get
$U_c\approx 10t$). In fact, to our knowledge, there exists no criterion that
yields a good estimate of $U_c$ in three dimensions.
\\  
In contrast to 3D where the effects of the thermal pairing fluctuations are
rather weak \cite{babaev,preosti}, in 2D they are very important
\cite{schneider}  leading apparently to a $T^*$ joining \emph{smoothly} $T_c$
\cite{singer}.  This confirms the observation by Moukouri {\it et al.}
\cite{moukouri} that precursor phenomena in the AHM have two origins: enhanced
thermal pairing fluctuations (in 2D only) and a strong pairing interaction (in
both cases). The fact that the AHM contains a transition between a Fermi liquid
and a state of ``incoherent pairs'' may be of interest in the context of the 
high-$T_c$ superconductors phase diagram, where the scenario of a hidden
quantum phase transition has been proposed \cite{loram}. Of course the driving
parameter in this case is the doping and the symmetry of the superconducting
phase is $d$-wave. 
\begin{acknowledgments}
The numerical calculations presented in this work were performed on the Eridan
server of the Ecole Polytechnique F\'ed\'erale de Lausanne (EPFL). This work
was supported by the Swiss National Science Foundation, IRRMA project, 
the University of Fribourg and the University of Neuch\^atel. We thank
F. Gebhard, F. Assaad, and T. Schneider for interesting discussions.
\end{acknowledgments}



\begin{thebibliography}{10}
\expandafter\ifx\csname bibnamefont\endcsname\relax
  \def\bibnamefont#1{#1}\fi
\expandafter\ifx\csname bibfnamefont\endcsname\relax
  \def\bibfnamefont#1{#1}\fi
\expandafter\ifx\csname url\endcsname\relax
  \def\url#1{\texttt{#1}}\fi
\expandafter\ifx\csname urlprefix\endcsname\relax\def\urlprefix{URL }\fi
\providecommand{\bibinfo}[2]{#2}
\providecommand{\eprint}[2][]{\url{#2}}

\bibitem{legett}
\bibinfo{author}{\bibfnamefont{A.~J.} \bibnamefont{Legett}}, in
  \emph{\bibinfo{booktitle}{Modern Trends in the Theory of Condensed Matter}},
  edited by \bibinfo{editor}{\bibfnamefont{A.}~\bibnamefont{Pekalski}}
  \bibnamefont{and} \bibinfo{editor}{\bibfnamefont{R.}~\bibnamefont{Przystawa}}
  (\bibinfo{publisher}{Springer Verlag}, \bibinfo{year}{1985}).

\bibitem{nsr}
\bibinfo{author}{\bibfnamefont{P.}~\bibnamefont{Nozi\`ere}} \bibnamefont{and}
  \bibinfo{author}{\bibfnamefont{S.}~\bibnamefont{Schmitt-Rink}},
  \bibinfo{journal}{J. Low Temp. Phys.} \textbf{\bibinfo{volume}{59}},
  \bibinfo{pages}{195} (\bibinfo{year}{1985}).

\bibitem{randeria}
\bibinfo{author}{\bibfnamefont{M.}~\bibnamefont{Randeria}}, in
  \emph{\bibinfo{booktitle}{Bose-Einstein Condensation}}, edited by
  \bibinfo{editor}{\bibfnamefont{A.}~\bibnamefont{Griffin}},
  \bibinfo{editor}{\bibfnamefont{D.}~\bibnamefont{Snoke}}, \bibnamefont{and}
  \bibinfo{editor}{\bibfnamefont{S.}~\bibnamefont{Stringari}}
  (\bibinfo{publisher}{Cambridge University Press}, \bibinfo{year}{1994}).

\bibitem{ranninger}
\bibinfo{author}{\bibfnamefont{R.}~\bibnamefont{Micnas}},
  \bibinfo{author}{\bibfnamefont{J.}~\bibnamefont{Ranninger}},
  \bibnamefont{and}
  \bibinfo{author}{\bibfnamefont{S.}~\bibnamefont{Robaszkiewicz}},
  \bibinfo{journal}{Rev. Mod. Phys.} \textbf{\bibinfo{volume}{62}},
  \bibinfo{pages}{113} (\bibinfo{year}{1990}).

\bibitem{hirsch}
\bibinfo{author}{\bibfnamefont{J.~E.} \bibnamefont{Hirsch}},
  \bibinfo{journal}{Phys. Rev. B} \textbf{\bibinfo{volume}{28}},
  \bibinfo{pages}{4059} (\bibinfo{year}{1983}).

\bibitem{lohgub}
\bibinfo{author}{\bibfnamefont{E.~Y.} \bibnamefont{Loh}} \bibnamefont{and}
  \bibinfo{author}{\bibfnamefont{J.~E.} \bibnamefont{Gubernatis}}, in
  \emph{\bibinfo{booktitle}{Electronic Phase Transitions}}, edited by
  \bibinfo{editor}{\bibfnamefont{W.}~\bibnamefont{Hanke}} \bibnamefont{and}
  \bibinfo{editor}{\bibfnamefont{Y.~V.} \bibnamefont{Kopaev}}
  (\bibinfo{publisher}{Elsevier Science Publishers}, \bibinfo{year}{1992}).

\bibitem{keller}
\bibinfo{author}{\bibfnamefont{M.}~\bibnamefont{Keller}},
  \bibinfo{author}{\bibfnamefont{W.}~\bibnamefont{Metzner}}, \bibnamefont{and}
  \bibinfo{author}{\bibfnamefont{U.}~\bibnamefont{Schollwock}},
  \bibinfo{journal}{Phys. Rev. Lett.} \textbf{\bibinfo{volume}{86}},
  \bibinfo{pages}{4612} (\bibinfo{year}{2001}).

\bibitem{capone}
\bibinfo{author}{\bibfnamefont{M.}~\bibnamefont{Capone}},
  \bibinfo{author}{\bibfnamefont{C.}~\bibnamefont{Castellani}},
  \bibnamefont{and} \bibinfo{author}{\bibfnamefont{M.}~\bibnamefont{Grilli}},
  \bibinfo{journal}{Phys. Rev. Lett.} \textbf{\bibinfo{volume}{88}},
  \bibinfo{pages}{126403} (\bibinfo{year}{2002}).

\bibitem{staudt}
\bibinfo{author}{\bibfnamefont{R.}~\bibnamefont{Staudt}},
  \bibinfo{author}{\bibfnamefont{M.}~\bibnamefont{Dzierzawa}},
  \bibnamefont{and}
  \bibinfo{author}{\bibfnamefont{A.}~\bibnamefont{Muramatsu}},
  \bibinfo{journal}{Eur. Phys. J. B} \textbf{\bibinfo{volume}{17}},
  \bibinfo{pages}{411} (\bibinfo{year}{2000}).

\bibitem{fakher}
\bibinfo{author}{\bibfnamefont{F.~F.} \bibnamefont{Assaad}},
  \bibinfo{author}{\bibfnamefont{W.}~\bibnamefont{Hanke}}, \bibnamefont{and}
  \bibinfo{author}{\bibfnamefont{D.~J.} \bibnamefont{Scalapino}},
  \bibinfo{journal}{Phys. Rev. B} \textbf{\bibinfo{volume}{50}},
  \bibinfo{pages}{12835} (\bibinfo{year}{1994}).

\bibitem{thesis}
\bibinfo{author}{\bibfnamefont{A.}~\bibnamefont{Sewer}},
Ph.D. thesis, \bibinfo{school}{Universit\'e de
  Neuch\^atel} (\bibinfo{year}{2001}).

\bibitem{engelbrecht}
\bibinfo{author}{\bibfnamefont{J.~R.} \bibnamefont{Engelbrecht}}
  \bibnamefont{and} \bibinfo{author}{\bibfnamefont{H.}~\bibnamefont{Zhao}},
  \eprint{cond-mat/0110356}.

\bibitem{schneider}
\bibinfo{author}{\bibfnamefont{T.}~\bibnamefont{Schneider}} \bibnamefont{and}
  \bibinfo{author}{\bibfnamefont{J.~M.} \bibnamefont{Singer}},
  \emph{\bibinfo{title}{Phase transtion approach to high temperature
  superconcuctivity}} (\bibinfo{publisher}{Imperial College Press},
  \bibinfo{year}{2000}).

\bibitem{dos}
\bibinfo{author}{
It would be interesting to compare the DQMC data with
DMFT results obtained using a more realistic density of states, e.g.
corresonding to a 3D tight-binding system, rather than the elliptic one
used in \cite{keller}.
}.

\bibitem{dossantos}
\bibinfo{author}{\bibfnamefont{R.} \bibnamefont{dos Santos}},
  \bibinfo{journal}{Phys. Rev. B} \textbf{\bibinfo{volume}{50}},
  \bibinfo{pages}{635} (\bibinfo{year}{1994}).

\bibitem{tremblay}
\bibinfo{author}{\bibfnamefont{S.}~\bibnamefont{Allen}},
  \bibinfo{author}{\bibfnamefont{H.}~\bibnamefont{Touchette}},
  \bibinfo{author}{\bibfnamefont{S.}~\bibnamefont{Moukouri}},
  \bibinfo{author}{\bibfnamefont{Y.~M.} \bibnamefont{Vilk}}, \bibnamefont{and}
  \bibinfo{author}{\bibfnamefont{A.~M.~S.} \bibnamefont{Tremblay}},
  \bibinfo{journal}{Phys. Rev. Lett.} \textbf{\bibinfo{volume}{83}},
  \bibinfo{pages}{4128} (\bibinfo{year}{1999}).

\bibitem{babaev}
\bibinfo{author}{\bibfnamefont{E.}~\bibnamefont{Babaev}},
  \bibinfo{journal}{Phys. Rev. B} \textbf{\bibinfo{volume}{63}},
  \bibinfo{pages}{184514} (\bibinfo{year}{2001}).

\bibitem{preosti}
\bibinfo{author}{\bibfnamefont{G.}~\bibnamefont{Preosti}},
  \bibinfo{author}{\bibfnamefont{Y.~M.} \bibnamefont{Vilk}}, \bibnamefont{and}
  \bibinfo{author}{\bibfnamefont{M.~R.} \bibnamefont{Norman}},
  \bibinfo{journal}{Phys. Rev. B} \textbf{\bibinfo{volume}{59}},
  \bibinfo{pages}{1474} (\bibinfo{year}{1999}).

\bibitem{singer}
\bibinfo{author}{\bibfnamefont{J.~M.} \bibnamefont{Singer}},
  \bibinfo{author}{\bibfnamefont{T.}~\bibnamefont{Schneider}},
  \bibnamefont{and} \bibinfo{author}{\bibfnamefont{M.~H.}
  \bibnamefont{Pedersen}}, \bibinfo{journal}{Eur. Phys. J. B}
  \textbf{\bibinfo{volume}{2}}, \bibinfo{pages}{17} (\bibinfo{year}{1998}).

\bibitem{moukouri}
\bibinfo{author}{\bibfnamefont{S.}~\bibnamefont{Moukouri}},
  \bibinfo{author}{\bibfnamefont{S.}~\bibnamefont{Allen}},
  \bibinfo{author}{\bibfnamefont{F.}~\bibnamefont{Lemay}},
  \bibinfo{author}{\bibfnamefont{B.}~\bibnamefont{Kyung}},
  \bibinfo{author}{\bibfnamefont{D.}~\bibnamefont{Poulin}},
  \bibinfo{author}{\bibfnamefont{Y.~M.} \bibnamefont{Vilk}}, \bibnamefont{and}
  \bibinfo{author}{\bibfnamefont{A.~M.~S.} \bibnamefont{Tremblay}},
  \bibinfo{journal}{Phys. Rev. B} \textbf{\bibinfo{volume}{61}},
  \bibinfo{pages}{7887} (\bibinfo{year}{2000}).

\bibitem{loram}
\bibinfo{author}{\bibfnamefont{J.~L.} \bibnamefont{Tallon}} \bibnamefont{and}
  \bibinfo{author}{\bibfnamefont{J.~W.} \bibnamefont{Loram}},
  \bibinfo{journal}{Physica C} \textbf{\bibinfo{volume}{318}},
  \bibinfo{pages}{194} (\bibinfo{year}{2001}).

\end{thebibliography}
\end{document}